\documentclass{aastex}
\usepackage{spr-astr-addons}

\begin{document}

\title{Polarization of AGN in UV Spectral Range.\\ 
 {\small (Accepted for publication in Astrophysics \& Space Science)}}

\author{Yu. N. Gnedin} \and \author{M. Yu. Piotrovich}
\and \author{T. M. Natsvlishvili}
\affil{Central Astronomical Observatory at Pulkovo,
Saint-Petersburg, Russia.} \email{gnedin@gao.spb.ru}

\begin{abstract}
We present the review of some new problems in cosmology and
physics of stars in connection with future launching of WSO. We
discuss three problems. UV observations of distant $z > 6$ quasars
allow to obtain information on the soft $< 1 KeV$ X-ray radiation
of the accretion disk around a supermassive black hole because of
its cosmological redshift. Really the region of X-ray radiation is
insufficiently investigated because of high galactic absorption.
In a result one will get important information on the reionization
zone of the Universe. Astronomers from ESO revealed the effect of
alignment of electric vectors of polarized QSOs. One of the
probable mechanism of such alignment is the conversion of QSO
radiation into low mass pseudoscalar particles (axions) in the
extragalactic magnetic field. These boson like particles have been
predicted by new SUSY particle physics theory. Since the
probability of such conversion is increasing namely in UV spectral
range one can expect the strong correlation between UV spectral
energy distribution of QSO radiation and polarimetric data in the
optical range. In the stellar physics one of the interesting
problems is the origin of the X-ray sources with super Eddington
luminosities. The results of UV observations of these X-ray
sources will allow to find the origin of these sources as
accreting intermediate mass black holes.
\end{abstract}


\section{Introduction}

World Space Observatory - Ultraviolet (WSO-UV) presents the
effective international project of a new space observatory
intended for the investigation of the Universe in the UV range of
the electromagnetic radiation spectrum
\citep{shustov09,shustov11}. The collection of scientific devices
of this Observatory allows to provide solutions to the most
important class of astronomical problems. It is very important that WSO-UV Observatory will realize the observations in the UV spectral region (100 $\div$ 320 nm) inaccessible for ground based instruments. In our work we present an observational program which can be successful in the frame of this cosmic project.

The first program presents the quasar observations at the high
cosmic distances $z = 5 \div 6$. These QSOs have been discovered
in the deep surveys \citep{fan03,jiang07,wang08,willot07}. UV
radiation in the $100 \div 320 nm$ range corresponds to the rest
frame radiation in the soft $h\nu \leq 0.5 KeV$ X-ray spectral
range. Namely this spectral region is inaccessible for the direct
observations of the remote Universe because of the strong
photoelectric absorption. The investigation of such remote QSOs
is very important for understanding the origin of the
reionization region that is located at the high cosmological
distances $z \geq 6$.

The second program is including the investigations of physical
processes at the intersection of cosmology and particle physics.
We analyze the physical processes of the interaction of UV photons
with intergalactic magnetic fields. One of the most curious recent
astronomical discovery is the cosmological alignment of the
electric vectors of polarization of distant QSOs
\citep{hutsemekers05,sluse05}. One of the most probable physical
mechanisms of this phenomenon is the process of the transformation
of the photon into axion in the magnetic field. Axion is predicted
by the modern supersymmetry particle physics (SUSY). The
probability of this process is increasing with photon energy and is especially high for UV photons. Therefore the observations of
QSOs with high polarized radiation namely in UV range allow to
obtain useful data about this physical process that is very
important especially for modern particle physics.

The third section of our paper is devoted to the problem of the origin of X-ray sources with super Eddington luminosities. The solution of this problem can be obtained by the observation of these sources namely in UV spectral region. We expect that in result of such observations the intermediate mass black holes will be revealed. These objects are now considered as the most probable candidates of ultra-luminous X-ray sources.

\begin{table}[t]
 \footnotesize
 \caption{Redshifts and continuum slopes.}
 \label{tab1}
 \begin{tabular}{@{}llr}
 \hline
 Quasar     & Redshift          & Slope$_{UV} (\alpha_{\nu})$\\
 \hline
 J0836+0054 & $5.810 \pm 0.003$ & $-0.62_{-0.06}^{+0.06}$\\
 J1030+0524 & $6.309 \pm 0.009$ & $ 0.46_{-0.25}^{+0.18}$\\
 J1044-0125 & $5.778 \pm 0.005$ & $-0.27_{-0.10}^{+0.09}$\\
 J1306+0356 & $6.016 \pm 0.005$ & $ 0.50_{-0.14}^{+0.12}$\\
 J1411+1217 & $5.927 \pm 0.004$ & $-0.21_{-0.19}^{+0.16}$\\
 \hline
\end{tabular}
\end{table}

\begin{table*}[t]
 \small
 \caption{Central BH masses ($10^9 M_{\odot}$)}
 \label{tab2}
 \begin{tabular}{@{}lccccc}
 \hline
 Quasar & $L_{bol}$ & $L_{bol} / L_{Edd}$ & $M_{BH}$ (C IV) & $M_{BH}$ (Mg II) & $M'_{BH}$ (Mg II)\\
 \hline
 J0836+0054 & 47.72 & 0.44 & $9.3 \pm 1.6$  & ...           & ... \\
 J1030+0524 & 47.37 & 0.50 & $3.6 \pm 0.9$  & $1.0 \pm 0.2$ & $2.1 \pm 0.4$ \\
 J1044-0125 & 47.63 & 0.31 & $10.5 \pm 1.6$ & ...           & ... \\
 J1306+0356 & 47.40 & 0.61 & $3.2 \pm 0.6$  & $1.1 \pm 0.1$ & $2.2 \pm 0.3$ \\
 J1411+1217 & 47.20 & 0.94 & $1.3 \pm 0.3$  & $0.6 \pm 0.1$ & $0.9 \pm 0.2$ \\
 J1623+3112 & 47.33 & 1.11 & ...            & $1.5 \pm 0.3$ & ... \\
 \hline
\end{tabular}
\end{table*}

\begin{table*}[t]
 \small
 \caption{Magnetic field of QSOs at the epoch of reionization.}
 \label{tab3}
 \begin{tabular}{@{}lcccrrr}
 \hline
 Quasar & z & $L_{bol} / L_{Edd}$ & $a_* = 0$,            &  $a_* = 0.95$,        & $ a_* = 0.998$,      & $a_* = 1.0$, \\
        &   &                     & $\varepsilon = 0.057$ &  $\varepsilon = 0.19$ & $\varepsilon = 0.32$ & $\varepsilon = 0.42$ \\
 \hline
 J0836+0054& 5.810& 0.44& 9.0x10$^{3}$ G& 7.5x10$^{3}$ G& 7.15x10$^{3}$ G& 6.6x10$^{3}$ G \\
 J1030+0524& 6.309& 0.50 & 1.5x10$^{4}$ G & 1.22x10$^{4}$ G & 1.16x10$^{4}$ G & 1.0x10$^{4}$ G \\
 J1044-0125& 5.778& 0.31& 7.1x10$^{3}$ G& 6.0x10$^{3}$ G& 5.7x10$^{3}$ G& 5.3x10$^{3}$ G \\
 J1306+0356& 6.016& 0.61& 1.8x10$^{4}$ G& 1.5x10$^{4}$ G& 1.43x10$^{4}$ G& 1.4x10$^{4}$ G \\
 J1411+1217& 5.927& 0.94& 3.5x10$^{4}$ G& 2.93x10$^{4}$ G& 2.8x10$^{4}$ G& 2.7x10$^{4}$ G \\
 J1623+3112& 6.247& 1.11& 3.5x10$^{4}$ G& 2.93x10$^{4}$ G& 2.8x10$^{4}$ G& 2.7x10$^{4}$ G \\
 \hline
\end{tabular}
\end{table*}

\section{QSOs at High Cosmological Distances: the Prospects of UV
Observations}

It is well known that the QSO radiation results from gas accretion on supermassive black holes (SMBHs). The SMBH is found in the central region of the active galactic nuclei (AGN). The typical values of SMBH masses are $10^6 \div 10^9 M_{\odot}$.

\subsection{The Study of the UV Slope}

The extremely high cosmological redshifts of a number of
extragalactic objects have been recently revealed. For QSOs these
redshifts are $z > 6$, for galaxies they are $z > 7$. The
gamma-ray bursts have been recently detected at $z = 8.2$. In deep surveys with the HST and ground based $8 \div 10$ m telescopes the galaxies and quasars at very large cosmological distances $z \sim 7 \div 10$ have been detected \citep{li02}. The data derived from observations are including cosmological redshifts, power-law indices of the spectral energy distribution, masses of quasars, their bolometric luminosities and the Eddington ratio, i.e. the ratio of the bolometric luminosity to the value of the Eddington luminosity. The quasars at $z \sim 5 \div 6$ have their masses at the level of $M_{BH} = 10^9 \div 10^{10} M_{\odot}$ and the Eddington ratio: $l_E = L_{bol} / L_{Edd} \sim 1$. Namely these objects are the prospective targets for observations at the cosmic observatory WSO-UV.

The spectral energy distribution in the rest frame of a typical
quasar is dominated by a broad optical-UV hump normally identified
with emission expected from an accretion disk surrounding a
massive black hole. Another most important component of quasar
radiation is X-ray emission that is originated in the close
vicinity of an accreting black hole. The effective power-law index
is $\alpha \sim 0.5$ in the optical region and in the UV. For high
redshift quasars index steepens further to $\alpha \sim 2$ in far
UV and soft X-ray region and seems consistent with extrapolation
to the soft X-ray excess detected in some quasars. Thus, the
observations with WSO-UV will allow to get the important information on spectral energy distribution of high redshift quasars in soft X-ray excess region.

Table 1 from \citet{jiang07} presents the redshifts and continuum
slopes of high redshift QSOs. The continuum slope of quasar rest
frame soft X-ray excess can be determined namely from UV
observation in our system. The central BH masses and the
bolometric luminosities are presented in Table 2 \citep{jiang07}.
The values of black hole masses are obtained from FWHM of CIV and
MgII. $M'_{BH}$ is estimated from the relation by
\citet{vestergaard06}.

\subsection{The Study of QSO Magnetic Fields}

It is now commonly accepted that AGNs and QSOs possess the
magnetized accretion disks. The magnetic coupling process is an
effective mechanism for transferring energy and angular momentum
from a rotating black hole to its surrounding accretion disk
\citep{zhang05}. Since the magnetic field on the horizon is
brought and held by accretion disk matter, some relation must exist between the magnetic field strength and the BH mass and the bolometric luminosity \citep{silantev09}:

\[
 B_H(G) = 6.2 \times 10^8
 \left(\frac{M_{\odot}}{M_{BH}}\right)^{1/2}
 \left(\frac{L_{bol}}{\varepsilon L_{Edd}}\right)^{1/2} \times
\]
\begin{equation}
 \times \frac{1}{1 + \sqrt{1 - a_*^2}}
 \label{eq1}
\end{equation}

This equation is valid for the equipartition condition at the BH
horizon. In Eq.(\ref{eq1}) $a_*$ is the Kerr parameter, the
Eddington luminosity $L_{Edd} = 1.3\times 10^{38} (M_{BH} /
M_{\odot}) erg / s$ and $\varepsilon$ is the radiation efficiency
of the accretion process, i.e. $L_{bol} = \varepsilon \dot{M}
c^2$, where $\dot{M}$ is the accretion rate.

The results of our calculations of the BH magnetic field strengths
at the reionization epoch ($z \geq 6$) are presented at Table 3.
The values of black hole masses $M_{BH}$ are taken from Table 2
\citep{jiang07}.

The magnetic field strengths and their topology in an accretion
disk around the supermassive black hole can be determined by the
UV polarimetric observations with use of Silant'ev and Gnedin
method based onto the account of the Faraday rotation effect on
the photon mean free path in the electron scattering process
\citep{gnedin97,silantev09}.

\subsection{The Study of Dust Emission in High Redshift QSOs}

Another effect which can be tested by WSO-UV is connected with the
problem of existence of the dust and molecular torus around the
SMBH in AGNs. In the local Universe the extremely UV and soft
X-ray radiation are absorbed by this torus. The UV observations of
high redshift QSOs allow to confirm absence or existence of the
molecular torus at the early stages of the Universe evolution.

\section{Investigations of UV Background Radiation: Searching
UV Photons Produced by Axion Decay and Photon Conversion into the
Intergalactic Magnetic Field}

\begin{figure}[t]
\includegraphics[width=8cm]{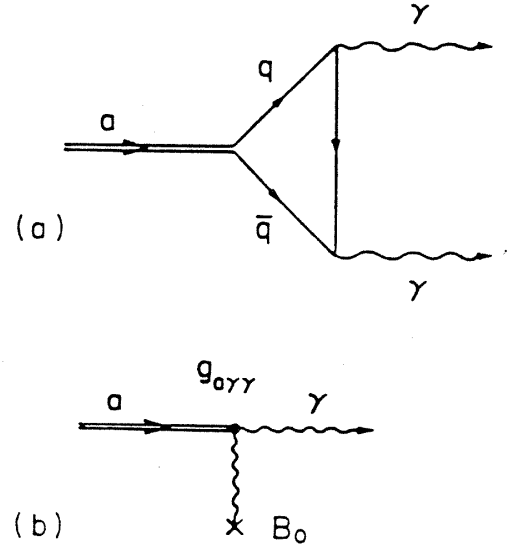}
\caption{(a) Two-photon coupling of the axion through a triangle
anomaly; (b) Axion-photon transition in a magnetic field.}
\end{figure}

Axion is the most popular candidate in the dark matter.  Though
the axion as a weakly interacting particle there exist two most
probable couplings of axions with the electromagnetic field
(Fig.1): the decay of axion into two photons and the process of
transformation (conversion) of axion into photon and the reverse
one of the photon conversion into axion.

Axions can decay into two photons. This $a \rightarrow 2 \gamma$
coupling arises due two different decay mechanisms: through
axion-pion mixing and via electromagnetic (EM) anomaly of PQ
symmetry in the standard QCD model. The axion decay time is
\citep{ressel91}:

\begin{equation}
 \tau_a (a \rightarrow 2 \gamma) \cong 6.8 \times 10^{24} \xi^{-2}
 \left(\frac{m_a}{1 eV}\right)^{-5} s
 \label{eq2}
\end{equation}

\noindent where $\xi = | E / N - 1.95 | / 0.72$, $E$ and $N$ are
the values of EM and color anomalies of PQ symmetry, respectively.

The free axion lifetime (\ref{eq2}) is sufficiently large to allow
for observations of decay of axions with mass much less than 1 eV.
However, axion interaction with magnetic fields can provide photon
production with energy comparable to the total axion energy (the
Primakoff effect).

The probability of axion conversion into a photon and the inverse
process of a photon conversion into an axion has the form
\citep{gnedin02}:

\begin{equation}
 P_{\parallel}(\gamma \leftrightarrow a) = \frac{1}{1 + x^2}
 \sin^2{\frac{1}{2} B_{\perp} g_{a\gamma} L \sqrt{1 + x^2}}
 \label{eq3}
\end{equation}

\noindent where $x = (\eta - 1) \omega / 2 B g_{a\gamma}$,
$\omega$ is the radiation frequency, $\eta$ is the dielectric
permittivity of the medium in that the radiation is propagating,
and $g_{a\gamma}$ is the constant of the interaction between axion
and photon fields. The conversion process is very sensitive to the polarization state of the photon, since only a single polarization state, when the electric vector oscillates into the plane of the directions of the magnetic field and propagation of the photon, is subject to the conversion. Here $B_{\perp} = B \sin{\theta}$ where $\theta$ is the angle between the photon propagation and the magnetic field. A commonly accepted system of units in (\ref{eq3}) is one for which $\hbar = c = 1$. For pure vacuum solution of Eq.(\ref{eq3}) transforms to

\begin{equation}
 P_{\parallel}(\gamma \leftrightarrow a) =
 \sin^2{\frac{1}{2} B_{\perp} g_{a\gamma} L} \approx
 B_{\perp}^2 L^2 g_{a\gamma}^2 / 4
 \label{eq4}
\end{equation}

\noindent in the case of a weak effect of conversion.

\begin{figure}[t]
\includegraphics[width=8cm]{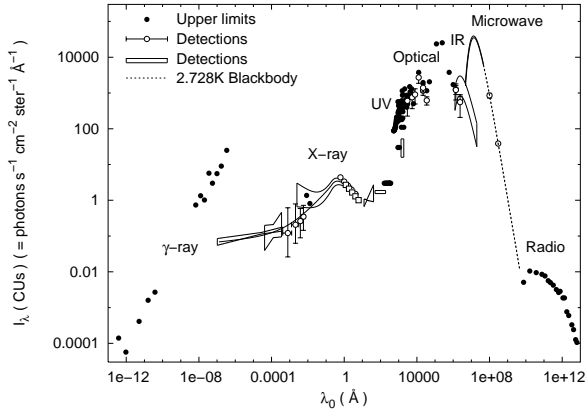}
\caption{A compilation of experimental measurements of the
intensity of cosmic background radiation at all wavelengths.}
\end{figure}

\begin{figure}[t]
\includegraphics[width=8cm]{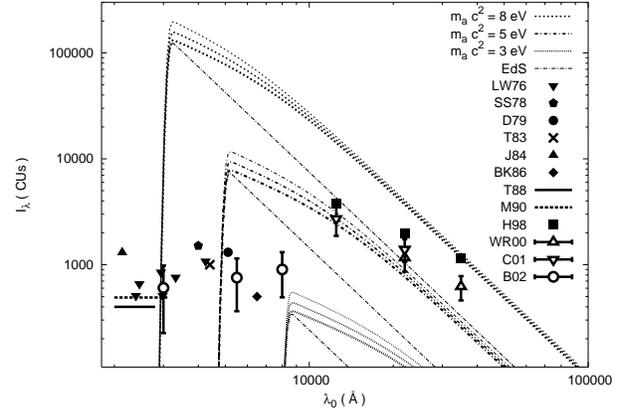}
\caption{The spectral intensity of the background radiation from
decaying axions as a function of observed wavelength.}
\end{figure}

The pseudoscalar axion like particles can contribute to observed
spectral energy distribution of extragalactic background
radiation. Fig.2 presents the spectral energy distribution of
observed background radiation of the Universe \citep{overduin04}.
Now the researches are conducted for detection of the contribution
of dark matter particles into the background photon spectrum. For
example, the decay and annihilation of massive dark matter
particles (WIMPs) can contribute in hard gamma-ray background
radiation. In UV region one can expect an additional contribution
to the background radiation of the Universe. We suggest that this
contribution can be provided namely by the process of axion
conversion into UV photon. The probability of this process
increases with photon energy because the value of $\eta - 1
\approx \omega_P^2 / \omega^2$, where $\omega_P^2 = 4 \pi e^2 N_e
/ m_e$ is the square of the plasma frequency, $N_e$ is the
electron density of the intergalactic medium. Fig.3 (from
\citet{overduin04}) presents the comparison of the observed
background radiation of the Universe with the expected contribution from the axion decay and transformation into intergalactic
magnetic field \citep{piotrovich09}.

\section{The Variability of Active Galactic Nuclei in the
UV Spectral Region.}

Many AGNs show variability of their radiation. As a rule, UV
variability scale is shorter that optical one \citep{lyuty05}. It
allows to investigate the well-known mechanism of the magnetic
reconnection and annihilation of the magnetic force lines in the
nearest region of supermassive black holes. The characteristic
time of this process is determined as $\Delta t = R_{\lambda} /
V_{rec}$, where $R_{\lambda}$ is the scale of an accretion disk
that corresponds to the radiation wavelength $\lambda$ and
$V_{rec} \sim V_A$ is the rate of reconnection process ($V_A$ is
the Alfven velocity). The scale of the standard accretion disk
\citep{shakura73} was calculated by \citet{poindexter07}:

\begin{equation}
 R_{\lambda} = 0.97 \times 10^{10} \lambda_{rest}^{4/3}
 \left(\frac{M_{BH}}{M_{\odot}}\right)^{2/3}
 \left(\frac{L_{bol}}{\varepsilon L_{Edd}}\right)^{1/3}
 \label{eq5}
\end{equation}

\noindent where $M_{BH}$ and $L_{bol}$ are the mass and bolometric
luminosity of a supermassive black hole, $\varepsilon$ is the
coefficient of radiative efficiency of accreting matter and
$L_{Edd} = 1.3 \times 10^{38} (M_{BH} / M_{\odot}) erg /s$ is the
Eddington luminosity, $\lambda_{rest}$ is the rest frame
wavelength in $\mu m$.

The relation between the typical variability scales in UV and
visual region is

\begin{equation}
 \frac{\Delta t (UV)}{\Delta t (Opt)} =
 \left[\frac{\lambda_{rest}(UV)}{\lambda_{rest}(Opt)}\right]^{4/3}
 \frac{V_{rec}(Opt)}{V_{rec}(UV)}
 \label{eq6}
\end{equation}

If we suggest that the reconnection velocity is proportional to the Alfven velocity and we use the standard model of an accretion disk, we obtain the reconnection time is proportional to $\lambda^2$.

Eq.(\ref{eq6}) means that one can derive the magnetic field
strength and its topology in the accretion disk in the case of
simultaneous measurement of characteristic variability time in the
UV and optical spectral regions, because the Alfven velocity is dependent essentially on the magnetic field.

\section{X-ray Sources with Supereddington Luminosity}

Recently a number of X-ray sources with extremal
supereddington luminosity has been discovered \citep{watson10}.
This fact allows to consider these sources as black holes of
intermediate masses. In this case it is easy to explain the
extremal luminosity of these objects.

The essential element of physical properties of these sources is
high ratio of X-ray flux to optical radiation $F_x / F_{opt} \sim
150\div 250$. This situation does not allow to make optical
identification of these sources. However in the UV spectral region
the situation is more optimistic because the ratio $F_x / F_{UV}
\sim 10$ if we use the standard model of an accretion disk. This
fact allows to expect the more favorable situation for
identification of these objects in UV spectral range. For
intermediate mass black holes the ratio of $F_x / F_{UV} \geq 1$, that is the decisive fact in the problem of identification of ultra luminous X-ray sources. Also another important task is to derive their variability scale and comparison of this scale with corresponding variability scales in X-ray and optical regions. In a result one should expect obtaining new important information on the origin of UV XRS.

\section{Conclusions}

Mentioned here tasks do not exhaust all possibilities of the
remarkable WSO-UV space observatory. One can expect that this
observatory provides the remarkable breakthrough in the
investigation of all of the Universe and separate astronomical
objects. One can expect the interesting results in new regions of close intersection of astronomy with fundamental particle physics. Many astrophysical objects can be considered as the particle physics laboratories. The origin of dark matter and dark energy is the problem which is very important for astronomy and fundamental physics. The solution of this problem may therefore lie in the existence of some new form of non-baryonic matter, such as, for example, axions.

The launching of WSO-UV Observatory will make essential
contributions to solution of these problems.

\acknowledgments

This research was supported by the program of Prezidium of RAS
''Origin and Evolution of Stars and Galaxies'', the program of the
Department of Physical Sciences of RAS ''Extended Objects in the
Universe'', by the Federal Program ''Scientific and
Science-Teaching Staff of Innovation Russia'' (GK 02.740.11.0246)
and by the grant from President of the Russian Federation "The
Basic Scientific Schools" NSh-3645.2010.2.

\end{document}